\documentclass[twocolumn,english,compsoc,journal]{IEEEtran}
\usepackage[unicode=true,
 bookmarks=true,bookmarksnumbered=true,bookmarksopen=true,bookmarksopenlevel=1,
 breaklinks=false,pdfborder={0 0 0},backref=false,colorlinks=false]
 {hyperref}
\usepackage[T1]{fontenc}
\usepackage{babel}

\hypersetup{pdftitle={A Self-Compiling Android Data Obfuscation Tool},
 pdfauthor={Alex Kolpa},
 pdfpagelayout=OneColumn, pdfnewwindow=true, pdfstartview=XYZ, plainpages=false}
\usepackage{breakurl}
\usepackage{color}
\usepackage{graphicx}
\usepackage{float}

\makeatletter

 \let\oldforeign@language\foreign@language
 \DeclareRobustCommand{\foreign@language}[1]{%
   \lowercase{\oldforeign@language{#1}}}

\usepackage[caption=false,font=normalsize,labelfont=sf,textfont=sf]{subfig}

\makeatother

\begin{document}

\title{A Self-Compiling Android Data Obfuscation Tool}

\author{Olivier Hokke, Alex Kolpa, Joris van den Oever, Alex Walterbos, and Johan Pouwelse (Course supervisor)}

\markboth{Delft University of Technology Student Project}{Your Name \MakeLowercase{\textit{et al.}}: Your Title}

\IEEEtitleabstractindextext{

\begin{abstract}
Smartphones are becoming more significant in storing and transferring data.
However, techniques ensuring this data is not compromised after a confiscation of the device are not readily available. 
DroidStealth is an open source Android application which combines data encryption and application obfuscation techniques to provide users with a way to securely hide content on their smartphones.
This includes hiding the application's default launch methods and providing methods such as dial-to-launch or invisible launch buttons. 
A novel technique provided by DroidStealth is the ability to transform its appearance to be able to hide in plain sight on devices. 
To achieve this, it uses self-compilation, without requiring any special permissions.
This Two-Layer protection aims to protect the user and its data from casual search in various situations.
\end{abstract}


}

\maketitle

\IEEEdisplaynontitleabstractindextext{}

\IEEEpeerreviewmaketitle{}

\section{Introduction}
\label{sec:introduction}
Using encryption technology on your smartphone may raise suspicion.
We present a tool which goes beyond encrypting sensitive data: It obfuscates the data, and itself.

There are over 1.5 billion smartphone users all over the world\cite{smartphoneUsage}, and over 80\% of the population in developed nations have mobile broadband subscriptions~\cite{mobileBroadband}.
A large percentage of this population captures their daily life on their phone.
For many, a large part of this data is personal and is something they want to keep private.

A large part of the Android security model is built around preventing attackers from unlocking the device. Once unlocked, either through forceful user cooperation or other means, a large part of the data becomes accessible. This gives a reason to actively encrypt and hide data from watchful eyes.

One such situation arose during the Arab Spring. 
It became clear that such sensitive data, stored on smartphones, can have a significant impact~\cite{arabSpring}.
Smartphone users provided on-the-ground information of the protests and government reactions through various social media platforms.
As such, these platforms get closely monitored or even blocked by some governments~\cite{cyberResponseGovernment}.
Because of this, spreading content directly from phone to phone has become a popular approach in places where conventional platforms are either locked down or under close surveillance.
However, this means that people carry sensitive material with them, posing the risk that the data will still be compromised. Governments such as Iran have even ordered citizens to hand over their devices and log in on their accounts~\cite{iranianCrackdown}.

It is obviously a necessity to be able to hide data until it can be shared in a safe manner.
In this paper we will explore data hiding on smartphones and present DroidStealth, a solution aiming to protect against casual smartphone searches.

\section{Problem Description}
\label{sec:problem-description}
As described, there is a need for a tool to \emph{hide} sensitive data, where encryption alone does not suffice.
To understand and properly describe the problem we have studied the variables involved, and defined them below.

\subsection{Audience}
The audience of the application is broadly defined.
DroidStealth's inspiration originates from the `citizen journalism'\cite{duffy2011} that played a significant role in the Arab Spring. 
Based on this, people dealing with an oppressive society was chosen as a worst case scenario.
However anyone interested in hiding and securing data on their device is part of the target audience for this application.

\subsection{Casual Search}
We aim to protect against \emph{casual search}.
In the scope of this article, we define `casual search' as follows: 
"A quick attempt to find information on a mobile device without applying advanced technical knowledge of the device, nor specific knowledge of the protecting methods proposed and implemented as described in this article".
Someone applying a casual search will be called a casual inspector, or casual searcher.
This reasoning also assumes that casual search is the most used search method:
A `quick glance' through a mobile device is more likely to occur than a thorough investigation of the device by someone who is experienced and trained where to look.
This would be time consuming and hard to manage on a larger scale.

\subsection{Beyond Simple Encryption}
An obvious solution to keep data safe from inspection by the wrong audience is encrypting the data.
Using a secure encryption has two useful properties:
First, the data would become inaccessible to any inspector.
Second, the encryption would render the files unreadable and unrecognizable to a human inspector, thus making them less incriminating.

However, finding inaccessible data can be a reason to extend a casual search with more intensive search methods, in which case the data might be compromised after all.
The same applies to the application that holds the data; any restriction to opening an application can be seen as suspicious.
To address both issues, the application should not only obfuscate the data, but the tool itself as well.

In this article, we will use the term `locked data' to refer to encrypted, inaccessible data.
With `locked application' we mean the application with its access restriction, as we will explain in Section~\ref{sec:approach}.

\subsection{Root Permissions}
We assume that users do not have \emph{root access} on their phones.
`Rooting' an Android device unlocks administrative privileges, allowing the user or applications to access restricted files, and communicate with device features more directly.
By doing this, it undermines Android's security model\cite{vidas2011all}.

It can be compared with root access as in Unix systems, and Administrator privileges in Windows systems.
Because `root access' allows manipulating system files, it provides more powerful tools that could benefit the solution to the problem addressed.

Enabling root access is a technical challenge, and has significant disadvantages for the user.
Therefore, DroidStealth does not require root access, and is therefore restricted from using its privileges.

\subsection{Spreading the application without a central app store}
\label{sec:problem-description:spreading}
In situations where a central distributor of applications, app store, is monitored and censored distribution of the application can prove difficult.
These stores can easily be inspected for installed applications, for which there is no way of circumventing, since it is the primary way to distribute these applications.
A simple search in Android's Play Store would be sufficient to expose the presence of the application, a risk many users would not be willing to take.
Not placing DroidStealth in the Play Store would prevent this problem, but makes distributing the application difficult.

A way of circumventing this limitation on distribution would be a `nomadic' distribution method for the application.
Instead of providing the application through a central point that can be monitored, the users should be able to share it directly with those interested.

\subsection{Data Type Restrictions}
The sensitive data that the user wants to keep safe can take many forms.
The data managed in DroidStealth will likely consist mostly of photos and videos, with the occasional document. 
However, we cannot predict exactly what data the user would want to hide in the application.
Therefore, the application should support the file types that the device could normally handle, since we do not want to restrict the user to certain file types.

\section{Approach}
\label{sec:approach}
Our solution combines password encryption and hiding of the DroidStealth tool itself.

\subsection{Password encryption}
The first layer of protection offered by DroidStealth is access restriction.
This restriction consists of two parts, one for the data itself and one for the application.

All data within DroidStealth is encrypted by default.
When the user adds a file to be managed by the application, it is automatically removed from its original location, encrypted, and added to a dedicated data folder.
This way, the data cannot be accessed by another application, and restricts users to opening these files exclusively through DroidStealth.
Even when the user wants to open an arbitrary file managed by DroidStealth, the conscious step of decrypting the file must be made first, raising awareness of the risk of exposure.

Every launch of the application requires the user to enter a pin code, which the user defines the first time DroidStealth is opened.
Only by entering this pin code upon launch can the user access the files managed by DroidStealth. Should it be forgotten, the data in the application will remain encrypted forever.

\subsection{Hiding of DroidStealth}
Finding encrypted and inaccessible data or applications can raise suspicion towards the user.
Since simply encrypting the data is not enough, our approach provides an added step of obfuscation that increases security of the data: DroidStealth hides itself.
This combination provides the two-layered protection, which is the key solution implemented in DroidStealth.

With the purpose of hiding all incriminating aspects of the encrypted data, we apply two ways of hiding DroidStealth on the device: 
Providing alternative launching methods, and `hiding the application in plain sight' by changing DroidStealth's appearance. 
The first is employed if the user wants to hide DroidStealth from easy access on Android devices, while changing its appearance allows for easy access as well as protection from casual search.

\subsubsection{Alternative Launch Methods}
The default way of launching an application on an Android device is through the `app drawer'.
This overview of all installed applications would normally show the DroidStealth logo, and provide a way of opening it.
The application would then be visible to anyone browsing the app drawer, including a casual inspector.

It is possible to hide the application from this drawer, but the user still needs to be able to launch DroidStealth when it is hidden.
For this, we have implemented two alternative launching methods, which can be enabled by the user in the application.

\textbf{Dialer Launch}
DroidStealth provides a launch option that opens the application by dialing a user-defined numeric sequence. 
The user enters this numeric sequence in the any regular phone number dialing application, as it would with a phone number.
Instead of actually calling the number, the application launches, requesting the pin code. 
Furthermore, DroidStealth fully intercepts the call, making sure the number never gets added to the call log.

\textbf{Transparent Widget}
Another option is to launch the app by means of an invisible widget on the device's home screen. 
This widget takes the form of a transparent area, thus showing the background image, positioned on the home screen by the user.
When the widget area is tapped once or twice, it remains inactive and simply appears to be an empty area.
However, when the widget is tapped five times consecutively, it launches DroidStealth.

\begin{figure}
\centering
\includegraphics[width=0.7\columnwidth]{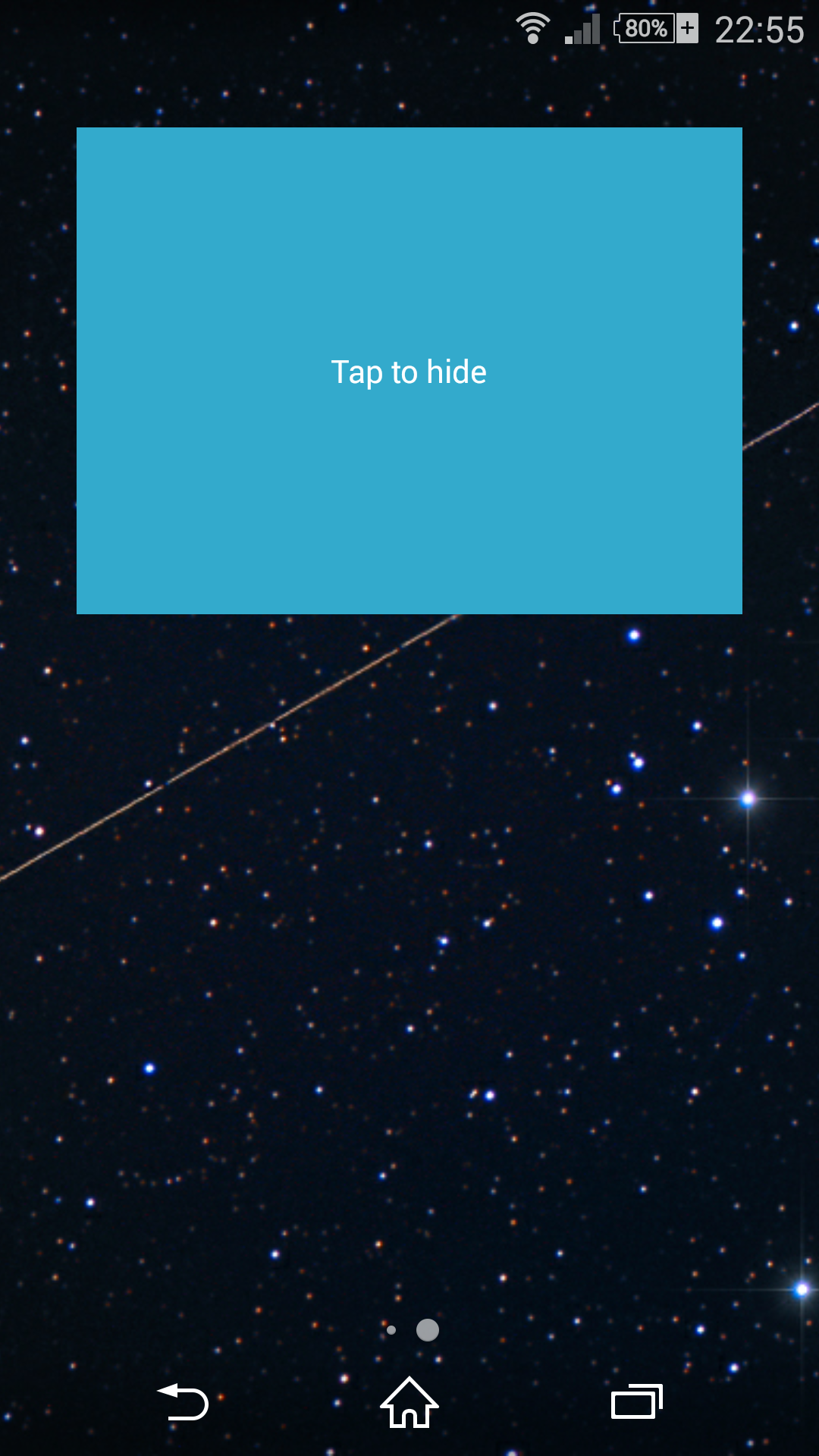}
\caption{A screenshot of the Android home screen containing a visible (top half) and an invisible (bottom half) DroidStealth Launch widget over a star-ridden background image.}
\label{fig:widget}
\end{figure}

When first created, this widget is temporarily visible, to help the user in placing it and to remember the location of the widget.
After the first time the widget is tapped, it becomes transparent.
A comparison of these two states of the widget is shown in figure~\ref{fig:widget}.

When adding a new access widget, all previously placed widgets become visible. 
This allows the user to retrieve forgotten widgets. 

\subsubsection{Morphing: Mimicking Other Applications}
DroidStealth is capable of transforming itself, changing the application's icon and name.
DroidStealth can then create an Android Application Package from which the application can be (re-)installed with the new appearance.
When composing its new appearance, DroidStealth provides all installed applications on the device. The user can therefore both supply its own name and icon, but it can also select an application it wants to mimic.

With the newly acquired appearance, DroidStealth does not need to be hidden from the app drawer and no secret launch methods are required.
This way, the intuitive `default' launching method is retained, while DroidStealth is still hidden from a casual inspector.

\subsection{Distribution}
As mentioned in Section~\ref{sec:problem-description:spreading}, having DroidStealth in a centralized store would result in a possible exposure threat.
For this reason, it was decided to let DroidStealth distribute itself nomadically. 
This does introduce some problems, most prominently the fact that users now need to enable `side-loading'.
Side-loading means enabling the installing of `untrustworthy' applications which came from outside the Android Play Store.
While this is a minor inconvenience, it was deemed less important that the visibility that comes from being in the Play Store.

\section{Implementation}
\label{sec:implementation}

This section diverges on the technical details of the implementation of DroidStealth.
It explains our solutions to challenges we encountered in the morphing and encryption.
Challenges and implementation decisions regarding the user experience are discussed, and finally an analysis of the vulnerabilities of DroidStealth is addressed.

\subsection{Morphing}
\label{sec:implementation:morphing}

As described in section \ref{sec:approach}, to hide the application some way is needed to alter the appearance of the application so that it can be hidden from casual search.
To achieve a complete change of appearance for an application in the app drawer, both its icon and its name need to be changed. 
However, to explain the full approach, some background on the inner workings of Android application packages might be required.

Android uses its own naming for its packages, so called Android Packages, or `.apk' files.
In reality, these files are very similar to Java archives -- so called `.jar' files -- namely that they are both archives containing binaries and resources.
To construct an apk file, one first needs the Dalvik binaries and the resources (images, animations and layouts among other things), which can then be compressed into an archive.
Optionally, the archive can be zipaligned, which improves its read performance by reordering the contents of the package as to optimize it for Android devices.
Once the archive has been constructed, it only needs to be signed with an appropriate key for the Android system to accept it as a valid package.

For the morphing to be successful, this process needed to be reversed, and then repeated after altering the archive resources.
For this, the original package is required. 
Fortunately, Android allows access to the original package from within the application without root access (see section \ref{sec:problem-description} for an explanation about what root access means).
Reversing of the package construction process, extraction of the files, can be achieved through normal zip extraction, something which is included by default in the Java version used for Android.
Once the files have been extracted, the application resources can be altered.
Due to the engineering effort of fully recompiling the binaries on an Android device, only support for initial self-compiling of the package was implemented. Further self-compilation, namely the decompilation, editing, and recompiling of the Dalvik binaries was considered beyond the scope of this work.

As explained earlier, both the application icon and the application name need to be altered to have a successful morphed application.
To alter the icons, first the original icon name that is stored in the resources is extracted from the application info, something which Android provides through its API.
Then it is a matter of iterating over the extracted content to find the appropriate icon files -- Android allows for multiple resolutions of the same image to be stored -- and replace them with the user specified icon.
Once the icons have been replaced, the name of the application needs to be changed.
Unfortunately, this is where the first restriction imposed by Android is encountered.

Android uses a resource map where each resource item gets mapped to a unique id generated by the compiler, which allows for easy re-use of resources in the making of an application.
This does pose an obstacle, since these reference maps are compiled into a binary format which is difficult to alter.
There is a solution available for desktop environments\cite{website:apktool}, but it relies on Android's Package Tool, `AAPT'.
Porting AAPT to Android proved to be near impossible due to its system requirements, which meant that decompiling the Android resources would not be a feasible approach.
Fortunately, the name of the application is accessed through a file known as the Android Manifest, which contains general information about the application package.
This manifest proved to be slightly more process-able from within the confinements of Android.
This meant that the new application name could be put directly into the manifest, without having to rely on the decompiling of Android resources.
After these two steps of altering the application appearance, the contents can be reconstructed again into a valid Android Package.

The first step is rebuilding the archive. 
Since it is structurally the same as a Java Archive, existing tools could be used. 
For this, the JarBuilder by Dominik Werthmueller\cite{website:jarbuilder} was used, since it posed the least amount of dependencies, which is favorable when working with Android, which can be rather picky about what parts of Java are actually supported in its system.
This resulted in a complete archive, which still needed to be zipaligned and signed with an appropriate key.
Because of similar restrictions posed by the decompiling of the Android resources, it was decided that including zip alignment was not achievable for now.

Finally, the package needs to be signed. 
Fortunately, an existing standalone solution is available outside the default Android signing methods, the `zip-signer' library\cite{website:zip-signer}.
However, a signing key still needs to be chosen. 
For the scope of this project, it was decided the test key would be sufficient, since actually signing it with appropriate keys which would needed to be tracked to prevent falsification of the application provide several challenges which will be discussed in section \ref{sec:limitations:morphing}.

Once the archive has been signed the morphing has been completed. 
The user can now be presented with an application package which contains the original application, albeit with a new appearance, according to the user's preferences.
This application can then be shared with other users.

\subsection{The Encryption Service}
\label{sec:implementation:encryption}
DroidStealth uses a `Service'\footnote{An Android Service is a part of an application that runs in the background, often used for more intensive or lengthy executions.} for the encryption of files.
This service runs in the background, and listens for \texttt{Intents} started by the application.
A queue is used to order the requests, and the service running in the backgrounds works through the queue continuously.

The encryption that DroidStealth provides is implemented using Facebook's Conceal API\cite{facebookConceal}.
Conceal provides a set of APIs for data encryption and authentication; we only use the first.
The Conceal library does not implement cryptography algorithms, but instead uses algorithms used in OpenSSL\cite{openssl}.

Files are encrypted individually.
When encryption would be applied at folder level, the application would have to decrypt all data upon being opened.
Not only does this expose all data during the user's interaction with (most likely) only one of the files;
but if the user would then forget to lock the data, all data managed by DroidStealth would remain exposed.
By applying per-file encryption, the risk of exposure is kept to a minimum.
Only loading files when needed also means an increase in startup time, as it is not necessary to decrypt the entire folder.

The process of encrypting an unencrypted file is quite simple because of the use of the Conceal API:
The \texttt{Crypto} class, provided by the Conceal API, handles all encryption logic in the process.
A `plain' Java input stream is created from the unencrypted file, and the \texttt{Crypto} class provides an output stream that encrypts data as it is being passed through.
When both streams have been created, a dedicated algorithm copies the data, buffered in chunks of 4096 bytes, from the unencrypted file's input stream to the encrypting output stream.
The final result is, as expected, the encrypted version of the previously unencrypted file.
The decryption process uses the exact opposite method:
An input stream provided by the \texttt{Crypto} class provides the decryption logic, by which the encrypted file is read.
The output of that stream is passed to a plain Java output stream, which writes the data to a new, unencrypted file.

The files are then stored in a folder outside the application folder.
To allow the updating of the application without data loss, this separation is required; overwriting the application folder may be required, especially when installing a morphed version of DroidStealth.
This means that the folder storing the files is public; other applications, as well as the user, can find it on their device via a computer or with a file explorer app.
The risk posed by this is migitated by the encryption of the files, as well as the requirement to know of DroidStealth, transcending the scope of `casual search'.

We see the above morphing procedure as the first minimal form of self-compilation for an application.

\subsection{User Experience}
\label{sec:user-experience}
Within this subsection we will elaborate on our design choices in user experience design related to locking and hiding of both data and the app.
We also briefly discuss the philosophy behind the styling.

\begin{figure}
\centering
\includegraphics[width=0.75\columnwidth]{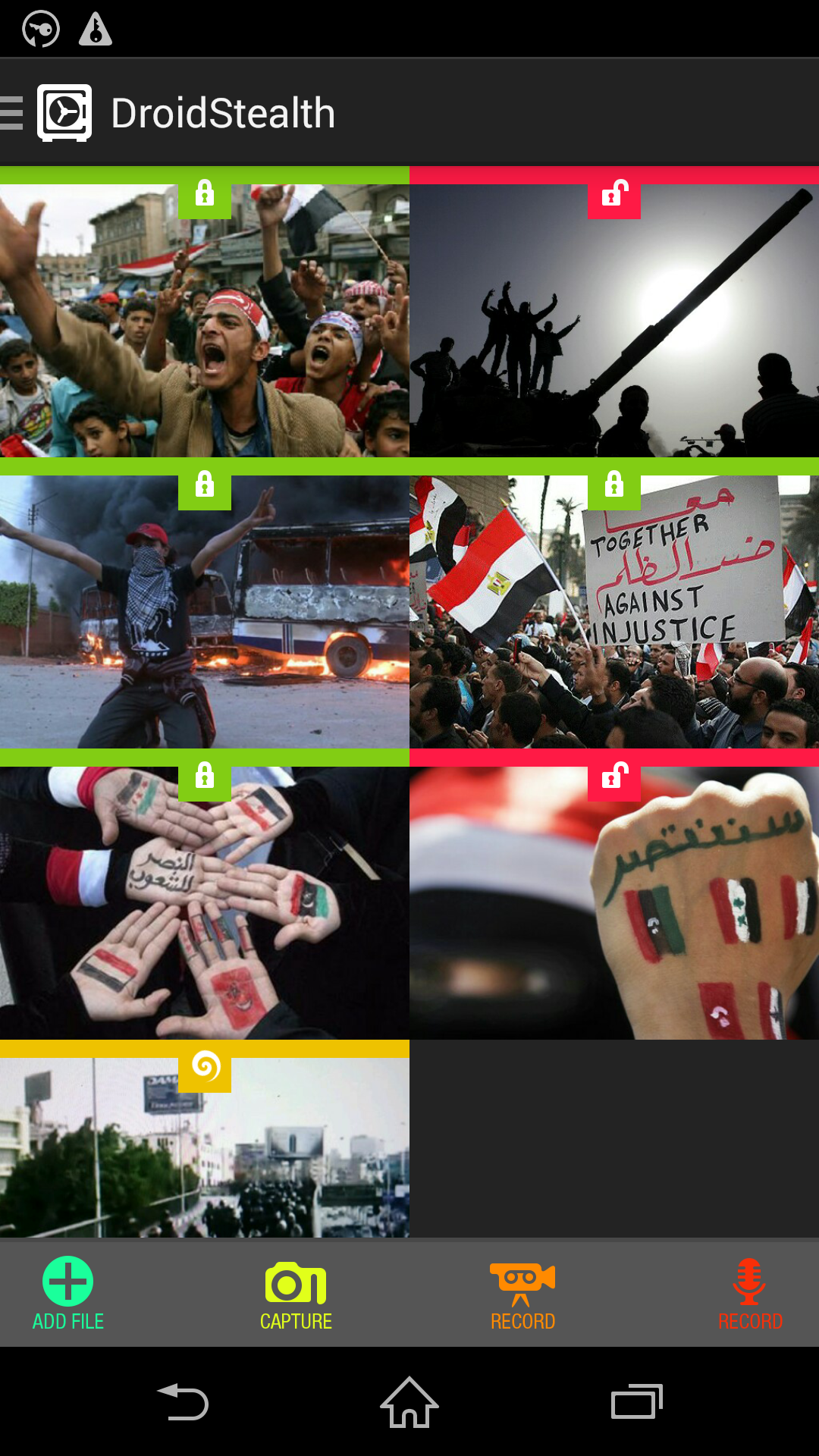}
\caption{The content gallery showing files in locked (green), unlocked (red) and currently processed (yellow) files. The upper-left corner shows the notification icons informing the user that files are being processed, and that other files are unlocked, respec}
\label{fig:gallery}
\end{figure}

Once the application has been opened, the user can interact with the files managed by DroidStealth.
Destroying an encrypted file is allowed, but opening and exporting/sharing files is not; the user will have to decrypt them first.
This extra step is part of the interaction to make the user aware that the files are then decrypted.

\begin{figure}[H]
	\centering
    \includegraphics[width=0.75\columnwidth]{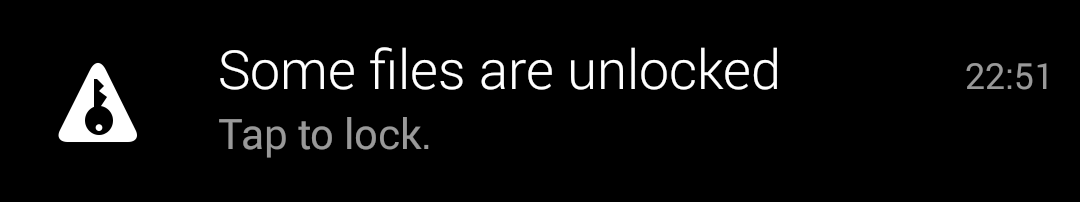}
    \caption{A non-disclosing notification to remind the user to lock accessible files.}
    \label{fig:notification}
\end{figure}

To add to this awareness, a warning is shown to the user, explaining that some files are unlocked and pose an exposure risk.
This warning is shown as a persistent notification in the notification drawer of the Android device (see Figure~\ref{fig:notification}.
When the notification is pressed, all unlocked files are immediately locked.
This provides a user friendly way to swiftly encrypt the files so they cannot be found, without the need to reopen the app.

Depending on the file size, encrypting or decrypting a file could take up to several minutes.
During this process, a notification which informs the user of the encryption or decryption process.
Once the user taps on it, the encryption or decryption is canceled by emptying the queue and finishing the current task.

\subsubsection{Launching DroidStealth}
\begin{figure}[H]
\centering
\includegraphics[width=0.6\columnwidth]{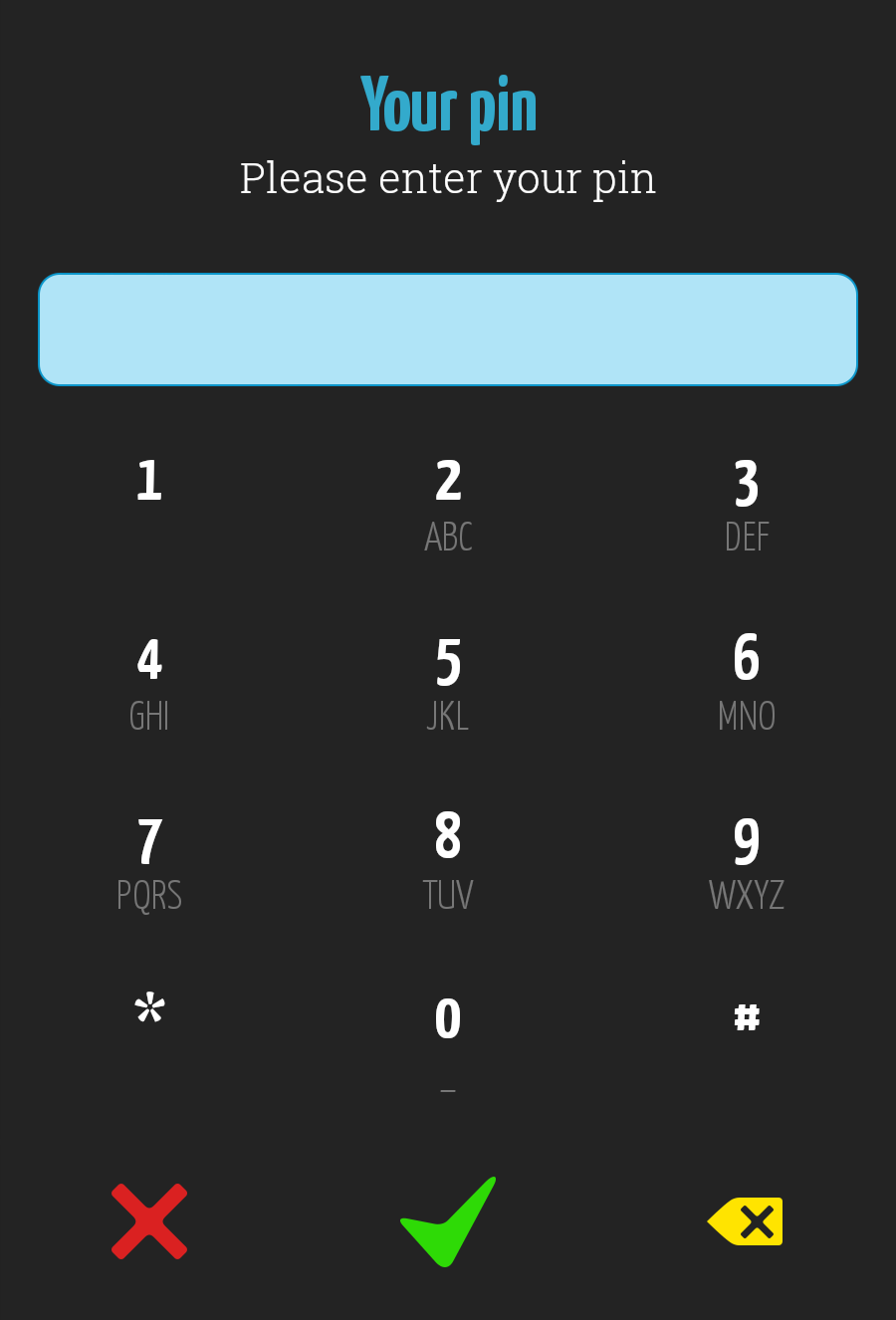}
\caption{The numerical keyboard presented to the user upon launching DroidStealth.}
\label{fig:keyboard}
\end{figure}

The trivial, default way of launching DroidStealth is through the app drawer of an Android device. 
When launching the application the first time, the user is prompted to enter a pin that will be used to access the application.
On following launches, the app will present the user with a numeric keyboard to enter its pin code (see Figure~\ref{fig:keyboard} .
If the pin is entered correctly, the DroidStealth will launch into the content gallery, as shown in Figure~\ref{fig:gallery}.
This pin code is required to open the application trough any launch method.
If the user forgets the pin code, the data in the application will remain encrypted forever.

\textbf{Alternative Launch Methods}
As explained in Section~\ref{sec:approach}, DroidStealth provides alternative launch methods.
The application provides a `Launch Menu' where launch methods can be enabled and disabled.
To protect the user from not being able to open the application, at least one launch method must be enabled at all times.

When the user adds a DroidStealth launch widget to the device's home screen, the new widget and all existing widgets are made temporarily visible.
This helps the user to place the new widget, as well as retrieve existing (lost) widget locations.
Touching any of the widgets will make all of them transparent again, and therefore invisible to the human eye.
From that point on, when the user presses the invisible widget 5 times, the app is launched, showing the pin keyboard.

\subsection{Styling}

DroidStealth is themed with a dark color in order to give users the feeling that they are indeed working in secret, and that their data will be safely hidden.
To give DroidStealth a professional ambience, we used a rounded but solid straight font for our titles, and a formal font for all other texts.
Combining this with the use of bright, but slightly softened primary and secondary colors, the user is provided with a balanced and consistent user interface.

The use of green and red is mirrored from the conventional meaning of those colors.
Green will always be used to indicate that something is good or safe, red will do the opposite.
Furthermore, we use the color orange when indicating a process is being executed.
Whenever a file is locking, or unlocking, the status bar of the file will be orange, and a animated twirl is shown to indicate that the file is being processed.

Finally, the gallery uses no padding or margins between the thumbnails of files, as we want to optimally make use of all the screen space.
Thumbnails are thus as big as they can be, and thus will be most efficient in showing the user its contents.

\subsection{Vulnerability Analysis}
\label{sec:vulnerability-analysis}
DroidStealth has some vulnerabilities, which are analyzed here.
Per vulnerability, we explain how it poses a risk, how big the risk is, and if applicable how this could be solved.

\subsubsection{Unlocked Files}
A potential security threat that can be exploited, is that
unlocked files are stored on the internal SD card of the device,
and that another app could constantly watch the folders in which
those files would be decrypted. Then this other app could save
them somewhere else or send them over the Internet.

Also, if files are unlocked, and the user's device is taken from
its rightful owner without warning, then the user might not have
gotten enough time to press the notifications to swiftly lock
the secret files back. Invaders won't be able to see the data
directly, unless they launch a file browser and start searching,
but they would be informed of the existence of secret files.
This could be a potential danger, because an attacker could be
interested in those files and do the user harm to get these
files. Users should be fully aware that they should only unlock
files if they are 100\% sure that they are in a safe location.

\subsubsection{Application visibility}
While it's possible to hide DroidStealth from the app drawer, or even change its appearance, there are still several ways its presence can be determined.
The most direct one would be to access the list of installed applications through the device settings. 
While it might take an attacker some time if DroidStealth is morphed, it would be possible to use this list to discover it through its package name, which doesn't change at this point.

More thorough search would make use of the `adb' tool, which can be used to communicate with authorized Android devices.
For this, enabling `USB Debugging' in Android's settings, physically connecting the device to a computer and allowing it access to the device is needed.
While this is be a possible scenario, it makes several changes to the security settings of the device (allowing external devices to access it), meaning it is considered out of the scope of `casual search' for this work.

\subsection{Spreading DroidStealth}
As expressed in the section \ref{sec:problem-description}, the application should be able to be spread nomadically.
This means there is no central distribution point, but users share the application between each other. 

Spreading the application is possible through a multitude of ways on Android, via all standard file sharing methods.
These methods include sending the application package through, for example, email, MMS, and BlueTooth.
DroidStealth also supports sharing via a more recently provided feature called Near Field Communication (NFC)\cite{website:nfc-spec}. 
When available on the device, it uses Android's NFC sharing feature called Android Beam to transfer the application to another device when it is held adjacent to the sending device.

These spreading methods allow the users to share DroidStealth directly from device to device, eliminating a central distribution point.

\subsection{Open Source}
This application was written as an open source project.
The code can be found on GitHub\cite{githubSource}, and may be used under terms of the licenses provided there.
The source code of this application is provided, as well as a separately usable library that can be used to include the morphing feature into another Android application.

\section{Conclusion and Future Work}
\label{sec:future-work}

DroidStealth beats casual search by combining encryption and the tool`s self-obfuscating approach.

There are several aspects of the application that can be improved. Which
can be categorized under one of the following: usability, and morphing.
The categories will be discussed individually as they are separated both
technically and conceptually. 

\subsection{Usability}
In terms of usability there has been no formal study on what aspects of the application work work well. 
However, informal testing has shown several avenues of improvement. 

The testing has shown that the alternative launch methods could use closer attention when it comes to the usability.
The counter-intuitive approaches to launching DroidStealth can be confusing.
There is also a risk that the methods, or the codes are forgotten when the user only sporadically opens the app;
This would render the data managed by the application lost forever.

With the providing of alternative access methods, easier access is provided to both the users and possible attackers.
The invisible widget that is provided is easily found in the widget list, though it is made more complex to detect when used in combination with a morphed DroidStealth.
The dialer launcher poses a risk when a user fills in the wrong launch code, as the entry won't be removed from the call log, and thus an attacked could be hinted towards the right pin. 
Also, if a user had made more mistakes, the attacker could merge the suspicious call log entries, and deduct the correct pin. 
A possible solution would be to check the call log for entries that are similar to the actual one, and to remove those as well.
Future work could address this issue with a solution that creates a more usable alternative launch method, while maintaining the security of the application.

Once the application has been launched, the user is required to pay close attention to the state of the application, and the state of the files managed by it.
A small human error can result in inadvertent data breaches, especially when the data in the application is opened in other apps.

Monitoring the application's life-cycle more closely could resolve some of these issues:
For example, `listening' to home button presses could be used to trigger the locking of files automatically whenever the app gets out of focus.
Another solution that has been discussed is limiting the time during which a file may be unlocked, which could restore the encryption after a (user-)designated duration.

\subsection{Morphing} 
\label{sec:limitations:morphing}

The basic features of the morphing library have been implemented, however it is not without limitations.
Two areas of improvement have been identified, namely app renaming restrictions and the lack of integrity verification.

Application names applied in the morphing process have to be the same length or shorter than the name originally given to the application.
Preliminary research indicates that this limitation can be overcome but not all details of packaged Android manifests have been reverse engineered.
Alternatively increased support for self-compilation could remove this limitation, as it would allow all details of the manifest to be changed as it would be done prior to compilation.

Integrity validation is not included in morphing.
With integrity validation we mean that the actual code itself is not modified to change the behaviour of DroidStealth, for example introducing a backdoor.
Normally this is something that the Android package system takes care of, but morphing has unique demands that makes it impossible to rely on built-in measures.
These measures rely on using a private key for signing the code, however to allow updating for the application after morphing the private key has to be bundled with the software so it can be signed again.
With access to this key anyone can create a modified version of the morphed application and spread that as an `update'.
Solving this problem is non-trivial and requires more research.

Other future work on the morphing, as mentioned in Section~\ref{sec:implementation:morphing}, would be a fully self-compiling system. Getting this to work would mean at least porting the `aapt' and `dex' tools provided by Google for construction of packages and binaries. Using existing systems on desktops, like ApkTool \cite{website:apktool}, should provide a good starting point for anyone interested in continuing this work.

\subsection{Re-installing the application}
When the application is re-installed from, for example, a morphed android application package, the encryption key used to lock data is replaced.
This means that the data locked before the re-installation is impossible to decrypt; a significant limitation in the usability.

A method of deriving a key that is consistent through re-installing the application is required, while it remains underivable for external parties.

\bibliographystyle{IEEEtran}
\bibliography{references}

\begin{thebibliography}{10}
\providecommand{\url}[1]{#1}
\csname url@samestyle\endcsname
\providecommand{\newblock}{\relax}
\providecommand{\bibinfo}[2]{#2}
\providecommand{\BIBentrySTDinterwordspacing}{\spaceskip=0pt\relax}
\providecommand{\BIBentryALTinterwordstretchfactor}{4}
\providecommand{\BIBentryALTinterwordspacing}{\spaceskip=\fontdimen2\font plus
\BIBentryALTinterwordstretchfactor\fontdimen3\font minus
  \fontdimen4\font\relax}
\providecommand{\BIBforeignlanguage}[2]{{%
\expandafter\ifx\csname l@#1\endcsname\relax
\typeout{** WARNING: IEEEtran.bst: No hyphenation pattern has been}%
\typeout{** loaded for the language `#1'. Using the pattern for}%
\typeout{** the default language instead.}%
\else
\language=\csname l@#1\endcsname
\fi
#2}}
\providecommand{\BIBdecl}{\relax}
\BIBdecl

\bibitem{smartphoneUsage}
\BIBentryALTinterwordspacing
``Smartphone users worldwide will total 1.75 billion in 2014.'' [Online].
  Available:
  \url{www.emarketer.com/Article/Smartphone-Users-Worldwide-Will-Total-175-Billion-2014/1010536}
\BIBentrySTDinterwordspacing

\bibitem{mobileBroadband}
\BIBentryALTinterwordspacing
``Statistics - international telecommunication union.'' [Online]. Available:
  \url{www.itu.int/en/ITU-D/Statistics/Pages/stat/default.aspx}
\BIBentrySTDinterwordspacing

\bibitem{arabSpring}
\BIBentryALTinterwordspacing
``Smartphones in the arab spring.'' [Online]. Available:
  \url{www.academia.edu/1911044/Smartphones_in_the_Arab_Spring}
\BIBentrySTDinterwordspacing

\bibitem{cyberResponseGovernment}
\BIBentryALTinterwordspacing
``Twitter revolutions and cyber crackdowns.'' [Online]. Available:
  \url{www.apc.org/en/system/files/AlexComninos_MobileInternet.pdf}
\BIBentrySTDinterwordspacing

\bibitem{iranianCrackdown}
\BIBentryALTinterwordspacing
``Iranian crackdown goes global.'' [Online]. Available:
  \url{http://www.wsj.com/articles/SB125978649644673331}
\BIBentrySTDinterwordspacing

\bibitem{duffy2011}
M.~J. Duffy, ``Smartphones in the arab spring,'' \emph{International Press
  Institute's 2011 report}, 2011.

\bibitem{vidas2011all}
T.~Vidas, D.~Votipka, and N.~Christin, ``All your droid are belong to us: A
  survey of current android attacks.'' in \emph{WOOT}, 2011, pp. 81--90.

\bibitem{website:apktool}
\BIBentryALTinterwordspacing
Brut.alll@gmail.com, ``Apktool.'' [Online]. Available:
  \url{https://code.google.com/p/android-apktool/}
\BIBentrySTDinterwordspacing

\bibitem{website:jarbuilder}
\BIBentryALTinterwordspacing
D.~Werthmueller, ``Jarbuilder.'' [Online]. Available:
  \url{http://sourceforge.net/projects/jarbuild/}
\BIBentrySTDinterwordspacing

\bibitem{website:zip-signer}
\BIBentryALTinterwordspacing
kellinwood, ``Zipsigner.'' [Online]. Available:
  \url{https://code.google.com/p/zip-signer/}
\BIBentrySTDinterwordspacing

\bibitem{facebookConceal}
\BIBentryALTinterwordspacing
``Facebook conceal, a cryptographics library.'' [Online]. Available:
  \url{https://facebook.github.io/conceal/}
\BIBentrySTDinterwordspacing

\bibitem{openssl}
\BIBentryALTinterwordspacing
``Openssl.'' [Online]. Available: \url{www.openssl.org}
\BIBentrySTDinterwordspacing

\bibitem{website:nfc-spec}
\BIBentryALTinterwordspacing
N.~Forum, ``Nfc forum technical specifications.'' [Online]. Available:
  \url{http://members.nfc-forum.org/specs/spec_list/}
\BIBentrySTDinterwordspacing

\bibitem{githubSource}
\BIBentryALTinterwordspacing
``Droidstealth and droidmorph source code.'' [Online]. Available:
  \url{https://github.com/droidstealth/droid-stealth}
\BIBentrySTDinterwordspacing

\end{thebibliography}

\end{document}